\documentclass[superscriptaddress,twocolumn,prb,showkeys,showpacs]{revtex4-2}

\usepackage[utf8]{inputenc}
\usepackage{amsmath}   
\usepackage{graphicx}   
\usepackage{verbatim}  
\usepackage{color}     
\usepackage{subfigure}  
\usepackage{hyperref}   
\usepackage{natbib}
\usepackage{gensymb}
\usepackage{color}
\usepackage{epstopdf}
\usepackage{multirow}
\usepackage{xfrac}

\begin{document}

\title{Conduction Band Structure and Ultrafast Dynamics of Ferroelectric  $\alpha$-GeTe(111)}

\author{Geoffroy Kremer}
\altaffiliation{Corresponding author.\\ geoffroy.kremer@univ-lorraine.fr}
\affiliation{D{\'e}partement de Physique and Fribourg Center for Nanomaterials, Universit{\'e} de Fribourg, CH-1700 Fribourg, Switzerland}
\affiliation{Institut Jean Lamour, UMR 7198, CNRS-Universit{\'e} de Lorraine, Campus ARTEM, 2 all{\'e}e Andr{\'e} Guinier, BP 50840, 54011 Nancy, France}

\author{Laurent Nicolaï}
\affiliation{New Technologies-Research Center University of West Bohemia, Plzen, Czech Republic}

\author{Frédéric Chassot}
\affiliation{D{\'e}partement de Physique and Fribourg Center for Nanomaterials, Universit{\'e} de Fribourg, CH-1700 Fribourg, Switzerland}

\author{Julian Maklar}
\affiliation{Fritz Haber Institute of the Max Planck Society, Faradayweg 4-6, 14195 Berlin, Germany}

\author{Christopher W. Nicholson}
\affiliation{D{\'e}partement de Physique and Fribourg Center for Nanomaterials, Universit{\'e} de Fribourg, CH-1700 Fribourg, Switzerland}
\affiliation{Fritz Haber Institute of the Max Planck Society, Faradayweg 4-6, 14195 Berlin, Germany}

\author{J. Hugo Dil}
\affiliation{Center for Photon Science, Paul Scherrer Institut, CH-5232 Villigen, Switzerland}
\affiliation{Institute of physics, Ecole Polytechnique F{\'e}d{\'e}rale de Lausanne, CH-1015 Lausanne, Switzerland}

\author{Juraj Krempask{\'y}}
\affiliation{Center for Photon Science, Paul Scherrer Institut, CH-5232 Villigen, Switzerland}

\author{Gunther Springholz}
\affiliation{Institut f{\"u}r Halbleiter-und Festk{\"o}rperphysik, Johannes Kepler Universit{\"a}t, A-4040 Linz, Austria}

\author{Ralph Ernstorfer}
\affiliation{Fritz Haber Institute of the Max Planck Society, Faradayweg 4-6, 14195 Berlin, Germany}
\affiliation{Institut f{\"u}r Physik und Astronomie, Technische Universit{\"a}t Berlin, 10623 Berlin, Germany}

\author{Jan Min{\'ar}}
\affiliation{New Technologies-Research Center University of West Bohemia, Plzen, Czech Republic}

\author{Laurenz Rettig}
\affiliation{Fritz Haber Institute of the Max Planck Society, Faradayweg 4-6, 14195 Berlin, Germany}

\author{Claude Monney}
\affiliation{D{\'e}partement de Physique and Fribourg Center for Nanomaterials, Universit{\'e} de Fribourg, CH-1700 Fribourg, Switzerland}

\begin{abstract}

$\alpha$-GeTe(111) is a non-centrosymmetric ferroelectric (FE) material for which a significative lattice distortion combined with a strong spin-orbit interaction gives rise to giant Rashba split states in the bulk and at the surface, which have been intensively probed in the occupied valence states using static angle-resolved photoemission spectroscopy (ARPES). Nevertheless, its unoccupied conduction band structure remains unexplored, in particular the experimental determination of its electronic band gap across momentum space. Using time-resolved ARPES based on high-repetition rate and extreme ultraviolet femtosecond (fs) laser, we unveil the band structure of $\alpha$-GeTe(111) in the full Brillouin zone, both in the valence and conduction states, as well as the exploration of its out-of-equilibrium dynamics. Our work confirms the semiconducting nature of $\alpha$-GeTe(111) with a 0.85 eV indirect band gap, which provides an upper limit for comparison to density functional theory calculations. We finally reveal the dominant scattering mechanisms of photoexcited carriers during the out-of-equilibrium dynamics under fs light pulses.

\end{abstract}
\date{\today}
\maketitle


\begin{center}
\textbf{I. INTRODUCTION}
\end{center}

\begin{figure*}[t]
\includegraphics[scale=0.245]{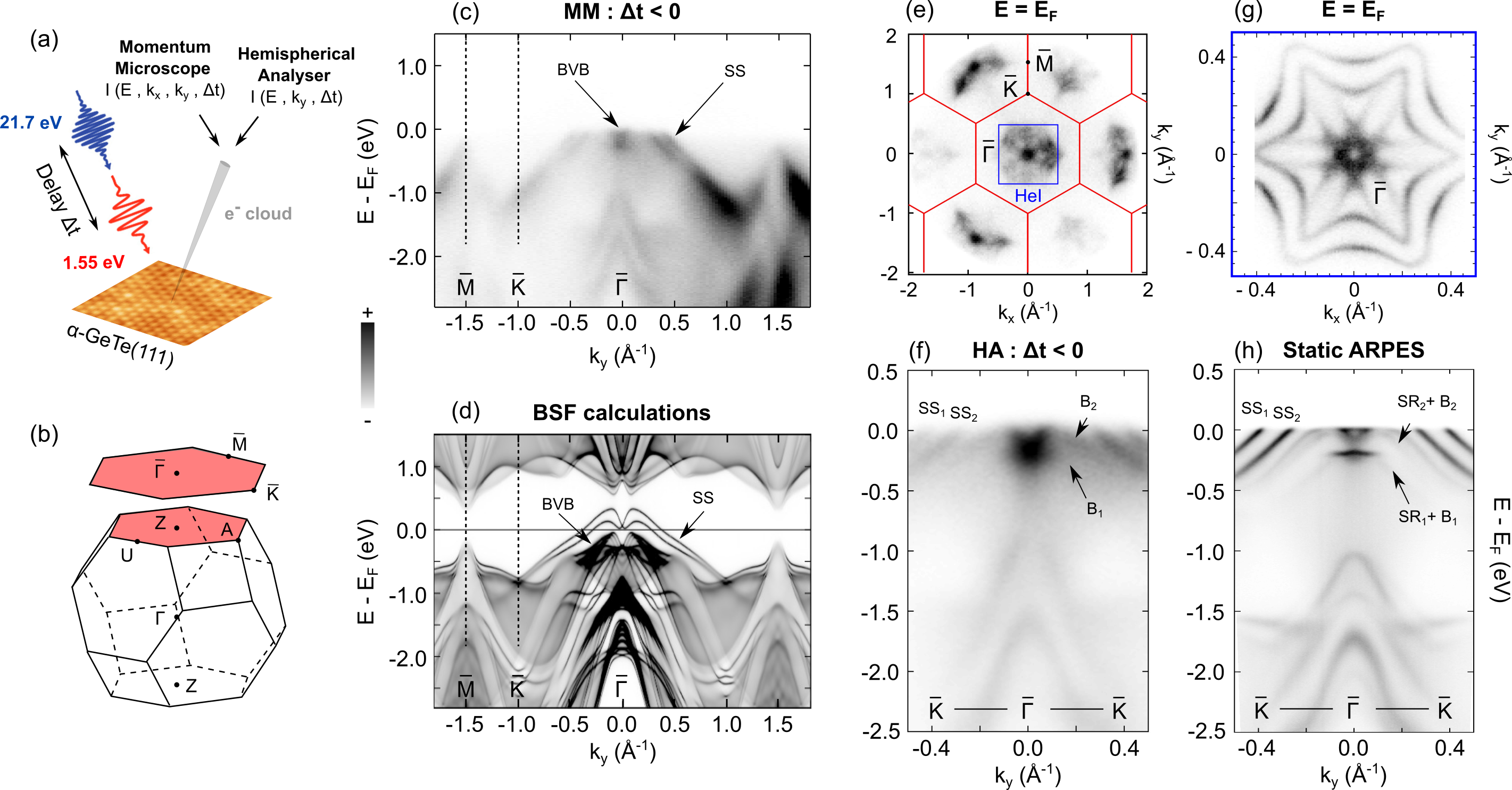}
\caption{\textbf{Valence band structure of $\alpha$-GeTe(111).} (a) Schematic tr-ARPES experiment with 21.7 eV probe (blue) and 1.55 eV pump (red) pulses. The emitted photoelectrons are collected as a function of their energy and momentum, either with an hemispherical analyser (HA) or with a momentum microscope (MM). (b) Three-dimensional Brillouin zone (BZ) of $\alpha$-GeTe(111) and its 2D projection in the (111) plane (red). (c) ARPES spectrum along the $\bar{M}-\bar{K}-\bar{\Gamma}-\bar{K}-\bar{M}$ high-symmetry direction obtained at a negative pump probe delay with MM and (d) corresponding Bloch spectral function (BSF) calculations for a Te-terminated surface of  $\alpha$-GeTe(111). (e) Constant energy surface taken at $E - E_{F} = 0$ meV obtained at a negative pump probe delay with MM. Red lines correspond to the hexagonal BZ obtained using the in-plane cell parameter of $\alpha$-GeTe(111) (a = 4.16 $\mathring{A}$). (f) ARPES spectrum along the $\bar{K}-\bar{\Gamma}-\bar{K}$ high-symmetry direction obtained at a negative pump probe delay with HA. All the tr-ARPES spectra have been acquired by integrating between -200 fs and -100 fs with respect to time zero. (g) High-resolution constant energy surface taken at $E - E_{F} = 0$ meV obtained with HeI$_{\alpha}$ radiation ($h \nu$ = 21.2 eV). The corresponding probed region is represented by the blue inset in panel (e). (h) Corresponding high-resolution ARPES spectrum along the $\bar{K}-\bar{\Gamma}-\bar{K}$ high-symmetry direction.}
\label{fig1}
\end{figure*}

Among the ferroelectric Rashba semiconductors (FERSC) family, \hbox{$\alpha$-GeTe(111)} is an appealing compound because it  exhibits a giant ferroelectric (FE) distortion below $T_{C} \approx 700 \thinspace K$, leading, in combination with a large spin-orbit coupling, to spin-polarized bulk and surface-split electronic states close to the Fermi level with the largest Rashba parameter so far reported \cite{di2013electric}. The spin-polarization of the bulk states can be reversed by flipping the FE polarization with the application of an external electric field \cite{di2013electric,krempasky2018operando}, which is promising in the perspective of low-power all-electric controlled
spintronic devices.  It has further a high potential in this context thanks to inverse spin Hall and inverse Rashba-Edelstein effects \cite{sanchez2013spin,lesne2016highly,noel2020non,varotto2021room} which are maximized in giant Rashba systems.  In that perspective, a lot of effort has been recently devoted to the characterization of the electronic band structure of \hbox{$\alpha$-GeTe(111)} in the occupied valence states and close to the Fermi level. A giant Rashba effect has been evidenced for bulk and surface states by angle-resolved photoemission spectroscopy (ARPES) \cite{di2013electric,liebmann2016giant,krempaskydisentangling,PhysResearchKrempasky2020,kremer2020,krempaskytriple}, as well as the associated spin polarization by spin-ARPES \cite{rinaldi2018ferroelectric,elmers2016spin}. A one to one relation between FE polarization and spin texture of the bulk states has been demonstrated, in agreement with state-of-the-art calculations \cite{rinaldi2018ferroelectric}. Recently, it has been shown that it is possible to maintain significant Rashba splitting down to a few monolayers \cite{croes2024pushing}, pushing down the thickness limit of FERSC. Nevertheless, only a few results have been obtained to explore the unoccupied states and the out-of-equilibrium dynamics which are of importance to obtain a complete understanding of the electronic band structure of \hbox{$\alpha$-GeTe(111)} and to understand transport properties in the perspectives of applications in devices \cite{kremer2022,clark2022ultrafast}. In particular, an experimental evidence of the nature and the magnitude of its electronic band gap remains elusive.  In the present work,  we investigate the ultrafast electron dynamics of
photoexcited \hbox{$\alpha$-GeTe(111)} at room temperature using time resolved (tr)-ARPES, allowing us to access both the valence and conduction band structure, as well as the microscopic scattering channels from the dynamics of the nonequilibrium state prepared by optical excitation. Combining a hemispherical analyzer (HA) and a time-of-flight momentum microscope (MM) for photoelectron detection, as well as extreme ultraviolet fs laser pulses obtained through high-harmonic generation (HHG) process, we are able to map the entire first Brillouin zone (BZ) both below and above the Fermi energy $E_{F}$\cite{puppin2019time,maklar2020quantitative,puppin2022excited}. Whereas the MM is adapted to large scale measurements, the HA allows detailed measurements close to high symmetry points. This combination is particularly relevant for \hbox{$\alpha$-GeTe(111)} which shows fine details in its band structure close to the center of the BZ that we resolve with the HA, and other electronic bands at the BZ edges that we reach with the MM. We confirm the semiconducting nature of $\alpha$-GeTe(111) in the bulk with a 0.85 eV indirect band gap and the presence of metallic Rashba split surface states crossing the Fermi level.


\begin{center}
\textbf{II. METHODS}
\end{center}

Five hundred nanometers-thick $\alpha$-GeTe(111) films were grown by molecular beam epitaxy (MBE) on BaF$_{2}$(111) substrate. The samples were transferred to the tr-ARPES setup using an ultra high vacuum (UHV) suitcase with a base pressure $< 1 \thinspace \times \thinspace 10^{-10}$ mbar. All tr-ARPES measurements were performed in UHV at a pressure $< 1 \thinspace \times \thinspace 10^{-10}$ mbar and at room temperature, using a laser-based high-harmonic-generation tr-ARPES setup \cite{puppin2019time} (\hbox{h$\nu_{probe}$=21.7 eV} with p polarization, h$\nu_{pump}$=1.55 eV with s polarization, \hbox{500 kHz} repetition rate, $\Delta E \thinspace \sim \thinspace $175 meV, $\Delta t\sim \thinspace $35 fs) with either a SPECS Phoibos 150 HA or a SPECS METIS 1000 time-of-flight MM, and a 6-axis manipulator (SPECS GmbH). The pump and probe spot sizes (FWHM) are $\sim \thinspace 150 \thinspace \times \thinspace 150 \thinspace \mu m^{2}$ and $\sim \thinspace 70 \thinspace \times \thinspace 60 \thinspace \mu m^{2}$, respectively. All discussed fluences refer to the absorbed fluence, determined using the complex refractive index\cite{shportko2008resonant} $ n = \sqrt{\epsilon} \sim 5.5 + 4.5 \thinspace i$ at $\lambda =800 $nm.  High-resolution ARPES measurements presented in Fig. \ref{fig1}(g,h) were performed on similar samples using a Scienta DA30 HA with monochromatised HeI$_{\alpha}$ radiation ($h \nu$ = 21.2 eV, SPECS UVLS with TMM 304 monochromator) at a temperature T = 50 K. Energy and angular resolutions were better than 10 meV and 0.1°.

\begin{figure*}[t]
\includegraphics[scale=0.235]{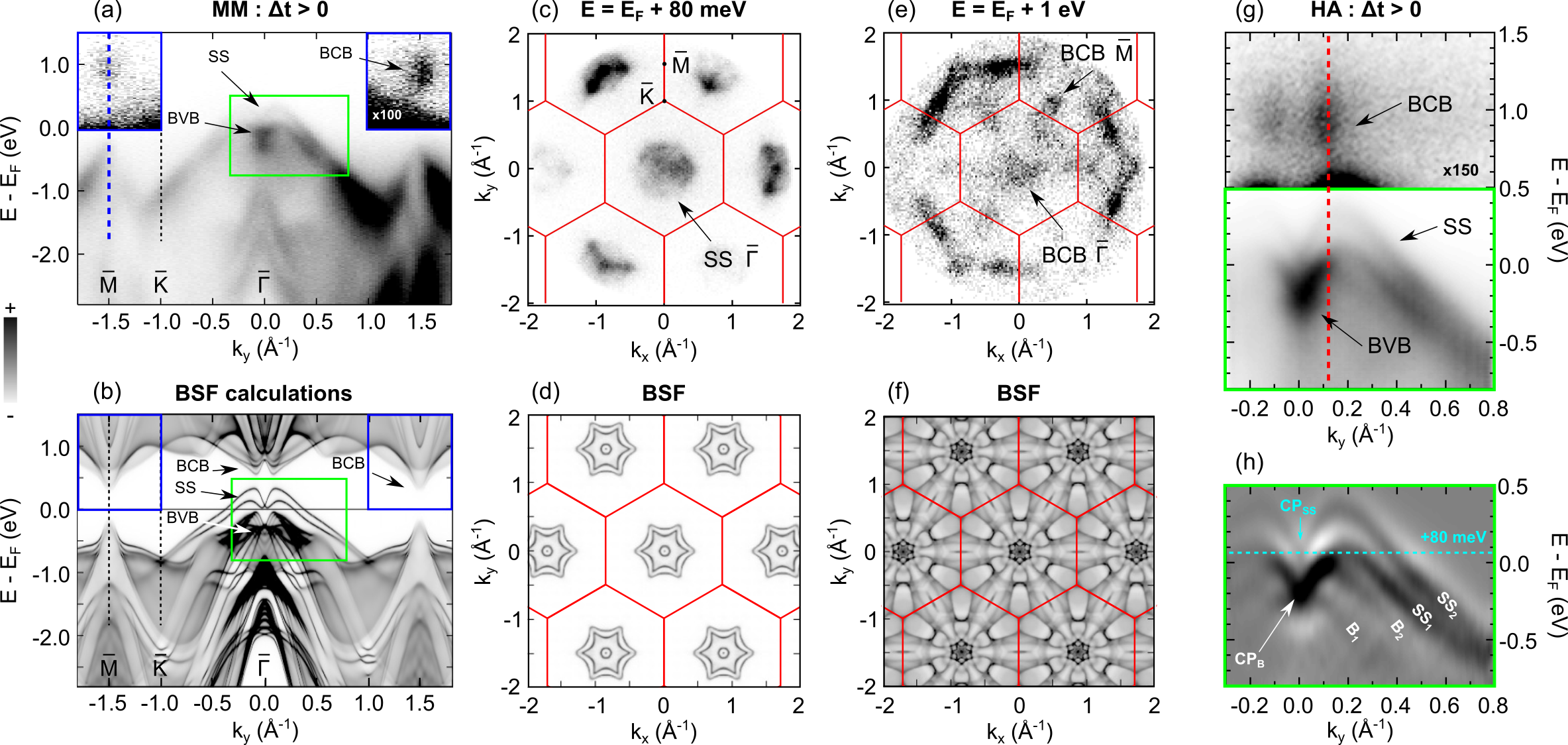}
\caption{\textbf{Conduction band structure of $\alpha$-GeTe(111).} (a) ARPES spectrum along the $\bar{M}-\bar{K}-\bar{\Gamma}-\bar{K}-\bar{M}$ high-symmetry direction obtained at a positive pump probe delay with MM (F =  0.35 mJ/cm$^{2}$) and (b) corresponding BSF calculations of \hbox{$\alpha$-GeTe(111).} (c,e) Constant energy surfaces taken at $E - E_{F} = +80$ meV and $E - E_{F} = +1$ eV with MM and (d,f) corresponding BSF calculations for a Te-terminated surface of  $\alpha$-GeTe(111). (g) ARPES spectrum along the $\bar{K}-\bar{\Gamma}-\bar{K}$ high-symmetry direction obtained at a positive pump probe delay with the HA (F =  0.75 mJ/cm$^{2}$). (h) Corresponding second derivative obtained from the green energy region of (g). All the tr-ARPES spectra have been acquired by integrating between +50 and +150 fs with respect to time zero. Blue and red dashed vertical lines in panels (a) and (g) are used for the discussion in Fig. \ref{fig3}.}
\label{fig2}
\end{figure*}

The presented ab-initio calculations are based on fully relativistic density functional theory as implemented within the multiple scattering Korringa-Kohn-Rostoker Green function based package (SPRKKR)\cite{ebert2011calculating}. Relativistic effects such as spin-orbit coupling are treated by Dirac equation. The local density approximation (LDA) was chosen to approximate the exchange-correlation part of the potential along with the atomic sphere approximation. The electronic structure is represented using the Bloch Spectral Function (BSF) which consists of the imaginary part of the Green function. A semi-infinite crystal of $\alpha$-GeTe(111) with Te surface termination was considered as in our previous work\cite{kremer2020,kremer2022}, representing both bulk and surface electronic states.


\begin{center}
\textbf{III. RESULTS}
\end{center} 


\textbf{Static investigation of the valence states.} The left part of Fig. \ref{fig1}(a) shows how we follow the out-of-equilibrium ultrafast dynamics of \hbox{$\alpha$-GeTe(111)} using a stroboscopic pump-probe approach. The principle is the following: we excite the material with infrared pulses ($1.55$ eV pump pulses) and examine the response of the system by probing its electronic band structure with the $21.7$ eV probe pulses after a given time delay $\Delta$t. At negative time delay, we probe the static valence states. By using the HA, we get a high precision both in terms of energy and momentum, whereas we access with the MM a large overview of the electronic dispersion in the full BZ.\cite{maklar2020quantitative}. This is particularly well illustrated in Fig.  \ref{fig1}(e) where the Fermi surface of \hbox{$\alpha$-GeTe(111)} recorded with the MM is shown. With this picture, we directly capture the periodicity of the electronic band structure over the first BZ, and up to the six second BZ neighbors. To better visualize this, the BZ of \hbox{$\alpha$-GeTe(111)} using the in-plane cell parameter  \hbox{(a = 4.16 $\mathring{A}$)} is superimposed in red, showing an excellent agreement with experiment. The band structure along the $\bar{\Gamma}-\bar{K}-\bar{M}$ high-symmetry direction (see the 2D BZ in Fig.  \ref{fig1}(b)) using the MM is presented in Fig.  \ref{fig1}(c), and compared to BSF calculations in Fig.  \ref{fig1}(d) performed for a Te terminated surface. In the vicinity of $E_{F}$ and close to the $\bar{\Gamma}$ point, the bulk and surface-derived states labeled BVB and $SS$ are well reproduced. This is also true for the electronic dispersions at other high symmetry points. For the $SS$, we clearly distinguish  a  momentum splitting which is better resolved by using the HA. Indeed, by focusing close to  $\bar{\Gamma}$ point in Fig.  \ref{fig1}(f), we now clearly resolve both the Rashba-split bulk state $B_{1}$ and $B_{2}$, and the Rashba-split surface state $SS_{1}$ and $SS_{2}$, in excellent agreement with both our calculations and the existing literature\cite{di2013electric,liebmann2016giant,krempasky2016disentangling,PhysResearchKrempasky2020,kremer2020,krempaskytriple,kremer2022}. It confirms that our surface is well ordered with Te termination, as evidenced by the surface state dispersion\cite{rinaldi2018ferroelectric}. It is further confirmed by concomitant high-resolution static ARPES measurements we performed on a similar sample but in a different experimental setup. These data are shown in Fig. \ref{fig1}(g,h) and exhibit state-of-the-art quality for this system\cite{PhysResearchKrempasky2020}, in particular through the resolution of bulk, surface and surface resonant (SR) states close to the Fermi level.

\begin{figure*}[t]
\includegraphics[scale=0.235]{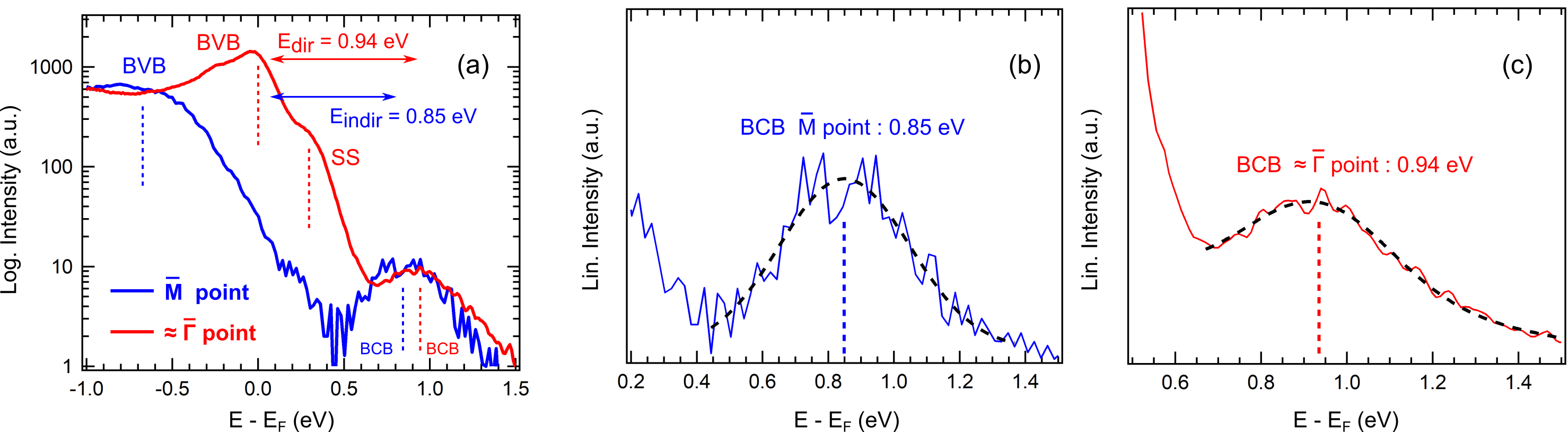}
\caption{\textbf{Estimation and nature of the electronic band gap.} (a) Energy distribution curves taken at the $\bar{M}$ and close to $\bar{\Gamma}$ (+ 0.11 $\mathring{A}^{-1}$) points (see blue and red dashed lines in  Fig. \ref{fig2}(a,g)) evidencing the direct and indirect band gap from different high-symmetry points in the BZ. The black dashed lines mark the best fits of the CB  (b) for the $\bar{M}$ point, and (c) close to the $\bar{\Gamma}$ point.}
\label{fig3}
\end{figure*}


\bigskip

\textbf{Exploration of the conduction states.} We now investigate the 
conduction states of \hbox{$\alpha$-GeTe(111)}.  It is a semiconducting material with a band gap $<$ 1 eV, so the pump photon-energy (1.55 eV) is sufficient to populate and experimentally observe the conduction band\cite{tsu1968optical}.  To do so, we probe the electronic band structure by integrating time delays over one hundred fs after time zero, \textit{i.e.} at positive $\Delta$t. Figure \ref{fig2}(a,b) displays the corresponding ARPES spectrum along the $\bar{\Gamma}-\bar{K}-\bar{M}$ high-symmetry direction using the MM and the corresponding BSF calculations, respectively. Experimentally, we can now distinguish some states above $E_{F}$, namely the dispersion of the Rashba splitted surface state $SS$ in the vicinity of the $\bar{\Gamma}$ point, and the bulk conduction band (BCB) at the $\bar{M}$ point of the BZ. This is again in excellent agreement with BSF calculations which show both these features (see green and blue rectangles for the correspondence). Further comparison between theory and ARPES is given in Fig. \ref{fig2}(c-f) with a full picture of the conduction states given by the MM at \hbox{$E = E_{F} + 80$ meV} and \hbox{$E = E_{F} + 1$ eV}. Whereas the first gives a picture of the Rashba splitted $SS$, the second highlights the BCB in the full BZ, showing its presence at both the $\bar{M}$ and $\bar{\Gamma}$ points at this particular energy, in good agreement with calculations. To get a better understanding of the dispersions close to the $\bar{\Gamma}$ point, we now turn to the HA measurements presented in Fig. \ref{fig2}(g,h). Panel (g) shows presented the raw ARPES data, exhibiting the dispersion of the bulk valence band (BVB) below  $E_{F}$, of the $SS$ below and above $E_{F}$, and the dispersion of the BCB close to \hbox{+1 eV} above $E_{F}$ at the center of the BZ. We do a more detailed analysis in panel (h) by plotting the second derivative (green rectangle: see panels (a,b,g)) of a region close to $E_{F}$. With this picture, we now clearly distinguish the $B_{1}$, $B_{2}$, $SS_{1}$ and $SS_{2}$ branches. We consequently resolve the full dispersion of the Rashba surface state and conclude that its crossing point at the $\bar{\Gamma}$ point (CP$_{SS}$) is localized at $80$ meV above $E_{F}$, in excellent agreement with our BSF calculations and our previous conclusions obtained by surface potassium doping\cite{kremer2020} (see Fig.  S1 in the Supplementary Material\cite{[{See Supplemental Material at }][{ for the comparison of bulk and surface electronic dispersions as obtained using static/time resolved ARPES and BSF calculations. It also includes additionnal bulk DFT calculations in the LDA approximation.}]supp}). Overall, this analysis is a demonstration of the powerful capabilities of the combination of the HA and MM with HHG to unveil the electronic dispersions of \hbox{$\alpha$-GeTe(111)} below and up to \hbox{+1 eV} above $E_{F}$, both in the full BZ and with a good energy/momentum resolution \cite{maklar2020quantitative}. It also confirms the excellent reliability of our theoretical model for this system, and the Rashba mechanism for both the bulk and surface derived states.

\bigskip


\textbf{Electronic band gap.} One of the most important physical property of semiconducting material is the electronic band gap. To know whether it is direct or indirect is an essential information that we can extract from tr-ARPES measurements in the low-excitation regime\cite{puppin2022excited} taking advantage of the energy and momentum resolution of the technique.  This has not been done so far for \hbox{$\alpha$-GeTe(111)}. As shown in Fig. \ref{fig2}(a) and Fig. \ref{fig2}(g), we resolve the BCB both at the $\bar{\Gamma}$ and at the $\bar{M}$ points of the 2D BZ, leading to the question  where its true minimum is located. We obtain this information by plotting energy distribution curves (EDCs) at both these high symmetry points, as illustrated in \hbox{Fig.  \ref{fig3}}. Whereas it is clear that the BVB maximum is located close to the $\bar{\Gamma}$ point at $E_{F}$, it is less obvious for the BCB minimum. Close to the $\bar{\Gamma}$ point, the BCB minimum is located at $E - E_{F}$ = +$0.94$ eV whereas at the $\bar{M}$ point, it is located at about $E - E_{F}$ = +$0.85$ eV, as extracted from a Gaussian fit of the CBM from EDCs (Fig.  \ref{fig3}(a-c)). So, the electronic band gap extracted from our tr-ARPES data equals +$0.94$ eV near the $\bar{\Gamma}$ point and +$0.85$ eV at the $\bar{M}$ point. It means that the electronic band gap of \hbox{$\alpha$-GeTe(111)} is indirect from close to the $\bar{\Gamma}$ point to the $\bar{M}$ point, in good agreement with our calculations and the literature \cite{di2013electric}.

Comparing these results to our calculations, the indirect nature of the band gap is predicted (see Fig. \ref{fig2}(b)) but the values are underestimated: 0.625 eV and 0.325 eV near the $\bar{\Gamma}$, and $\bar{M}$ points, respectively. They are performed in the framework of the LDA approximation, which is known to underestimate the band gap of semiconducting materials \cite{kremer2021dispersing}. This has  been shown in the literature for the case of \hbox{$\alpha$-GeTe(111)} by comparing DFT calculations using Perdew-Burke-Ernzerhof (PBE)  to the most accurate Heyd-Scuseria-Ernzerhof (HSE) hybrid functional\cite{di2013electric}. For sake of clarity, we confirm our conclusions regarding the bulk band gap by performing in Fig. S2 (a) complementary BSF calculations close to the Fermi energy in the LDA approximation including solely the bulk states, i.e. without the surface states contributions. The value extracted from tr-ARPES in the present work is close to the expected value from DFT calculation using the HSE approximation close to the $\bar{\Gamma}$ point (0.94 eV compared to 0.96 eV) whereas it is worse at the $\bar{M}$ point (0.85 eV compared to 0.68 eV)\cite{di2013electric}. This discrepancy between experiment and theory can likely be partially explained by a $k_{z}$ effect occurring due to the photoemission process. Assuming a free-electron-like final state assumption, the out-of-plane momentum $k_{z}$ of a photoelectron emitted is given by the relation $k_{z} = 0.512 \times \sqrt{E_{kin} \thinspace cos^{2} \theta + V_{0}}$. It is consequently not the same for an angle of extraction $ \theta = 0^{\circ}$ (close to the $\bar{\Gamma}$ point) and $ \theta = 44^{\circ}$ (at the $\bar{M}$ point). Assuming further a photoelectron kinetic energy of 18 eV and an inner potential of $V_{0} = 15$ eV, we obtain for $\bar{\Gamma}$ and $\bar{M}$ points, $k_{z} = 2.94 \thinspace \mathring{A}^{-1}$ and $k_{z} = 2.52 \thinspace \mathring{A}^{-1}$ , respectively\cite{liebmann2016giant}. Regarding the half periodicity of $\alpha$-GeTe(111) along $k_{z}$ in the 3D BZ ($|| \Gamma Z || = 0.93 \thinspace \mathring{A}^{-1}$) \cite{krempasky2016disentangling}, it means that the BCB at the $\bar{\Gamma}$ point is close to the (AZU) plane by \hbox{+ 0.15 $\mathring{A}^{-1}$} and the $\bar{M}$ point is located at \hbox{- 0.27 $\mathring{A}^{-1}$} from this plane. To be exhaustive, it means that our measurement near the $\bar{\Gamma}$ point corresponds to a probing at \hbox{+ 0.15 $\mathring{A}^{-1}$} from the Z point in the 3D BZ whereas at the $\bar{M}$ point it corresponds to a probing at \hbox{- 0.27 $\mathring{A}^{-1}$} from the L point: see the schematic green line in the extended BZ picture in Fig. S2 (b). This difference likely explains why the agreement between experiment and theory is better for the direct band gap value compared to the indirect one. Indeed, as evidenced in our DFT calculations in Fig. S2 (a), the bulk band gap shows a significant $k_{z}$ dependence. It increases when moving away from both L (to U point) and Z (to $\Gamma$ point) points. In summary, the 0.85 eV value of the indirect band gap extracted from tr-ARPES can serve as an upper limit for the indirect electronic band gap of $\alpha$-GeTe(111), as expected from DFT calculations (0.65 eV with HSE functional) and optical measurements (0.6 eV) in the literature\cite{di2013electric,tsu1968optical}. Our analysis consequently points out some difficulties of tr-ARPES to properly extract the exact value of the indirect band gap in a material where the BCB minimum and BVB maximum are located at distant points in the BZ, due to the photoemission process. To overcome this issue, a continuous tuning of the probe energy could be performed to probe the energy position of the BCB and BVB at high-symmetry points in the 3D BZ (in our case Z and L points), which is not possible at the moment in the UV range. Another possibility would be to perform tr-ARPES measurements at higher probe energy to reduce the angle of extraction of photoelectrons at higher momenta and consequently reduce the $k_{z}$ variation, at the cost of momentum resolution. Last but not least, it has been demonstrated that tr-ARPES can measure the excited band gap and not the true fundamental band gap of solids\cite{puppin2022excited}. In our measurements, we are using a low-excitation fluence as defined in ref [19] such that we do not expect significant deviation of the BCB position as a function of fluence, given the experimental energy resolution.

\bigskip


\begin{figure*}[t]
\includegraphics[scale=0.235]{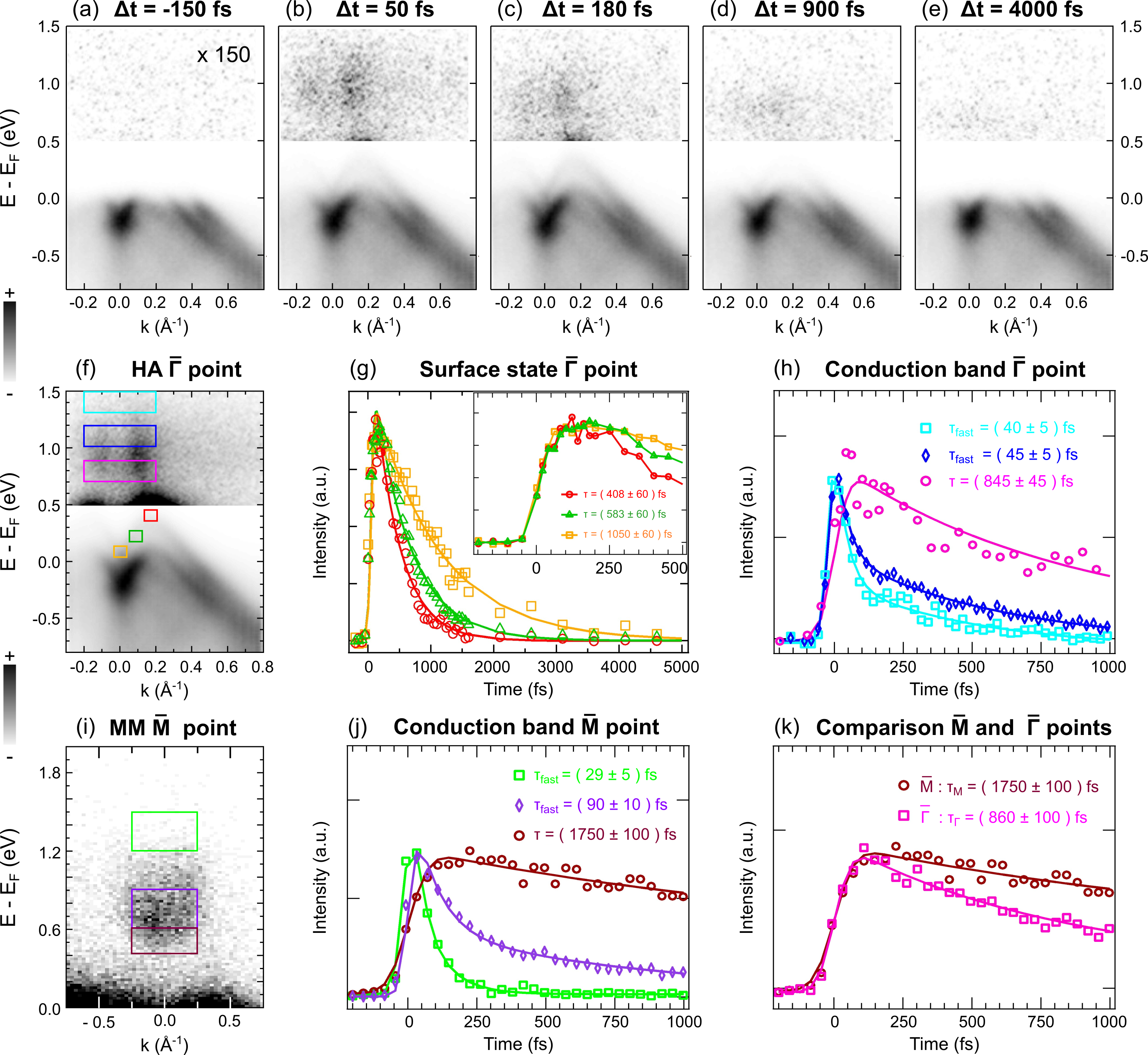}
\caption{\textbf{Transient dynamics of $\alpha$-GeTe(111) at the $\bar{\Gamma}$ and $\bar{M}$ points of the BZ.} (a-e) Series of ARPES spectra (HA) along the $\bar{K}-\bar{\Gamma}-\bar{K}$ high-symmetry direction obtained at different pump-probe delays from -150 fs to +4000 fs. (f) Time integrated ARPES spectrum close to the $\bar{\Gamma}$ point obtained with the HA (F =  0.5 mJ/cm$^{2}$) between +50 and +150 fs with respect to time zero. Colored boxes show the (g) transient electronic population at different energy/momentum of the Rashba-split surface state above the Fermi level (orange/green/red) and the same for (h)  the conduction band at the $\bar{\Gamma}$ point (pink/blue/cyan). (i) Time integrated ARPES spectrum close to the $\bar{M}$ point obtained with the MM (F =  0.35 mJ/cm$^{2}$) between +50 and +150 fs with respect to time zero. Colored boxes show the (j) transient electronic population at different locations in the conduction band at the $\bar{M}$ point (brown/purple/green), and the (k) comparison of the relaxation dynamics at the minimum of the conduction band at the $\bar{\Gamma}$ and $\bar{M}$ points using the MM. For panels (g,h,j,k), symbols are the experimental data points and lines the best fits obtained by using a single (double in the case of "fast" subscript, see discussion in the text) exponential relaxation procedure and giving the noticed relaxation times. Note that for each ARPES spectrum of panels (a-f), the photoemission intensity of the upper part has been multiplied by a factor 150 with respect to the lower part.}
\label{fig4}
\end{figure*}

\textbf{Out-of-equilibrium dynamics and scattering channels.} Let us now discuss the out-of-equilibrium dynamics of \hbox{$\alpha$-GeTe(111)}. Contrary to what we have achieved so far, we now no longer integrate over one hundred fs before or after time zero, but look at each ARPES spectrum at different time delay $\Delta$t. A corresponding series of ARPES spectra recorded at different time delays are shown in Fig. \ref{fig4}(a-e) close to the $\bar{\Gamma}$ point. At negative time delay ($\Delta t = - 150$ fs), only the valence states are visible. After $50$ fs, some electrons are populating the BCB over a broad energy and momentum range and the dispersion of the $SS$ above $E_{F}$ starts to be visible. At $\Delta t = 180$ fs, the electrons accumulate at the bottom of the BCB and $SS$'s intensity is more pronounced. Then, after $900$ fs the photoemission intensity in both regions progressively decreases and the system is back to the equilibrium state after $4$ ps. This dynamics is typical of semiconducting material where the carriers relax their excess energy  by scattering events which transiently populate bands at different momenta and finally accumulate at the CB bottom. Here, the carriers dwell for longer time at the BCB minimum because of the band gap: electrons lose their energy by recombining with holes in the BVB, a process that can take a few ps to occur \cite{bertoni2016generation}.

\begin{figure*}[t]
\includegraphics[scale=0.63]{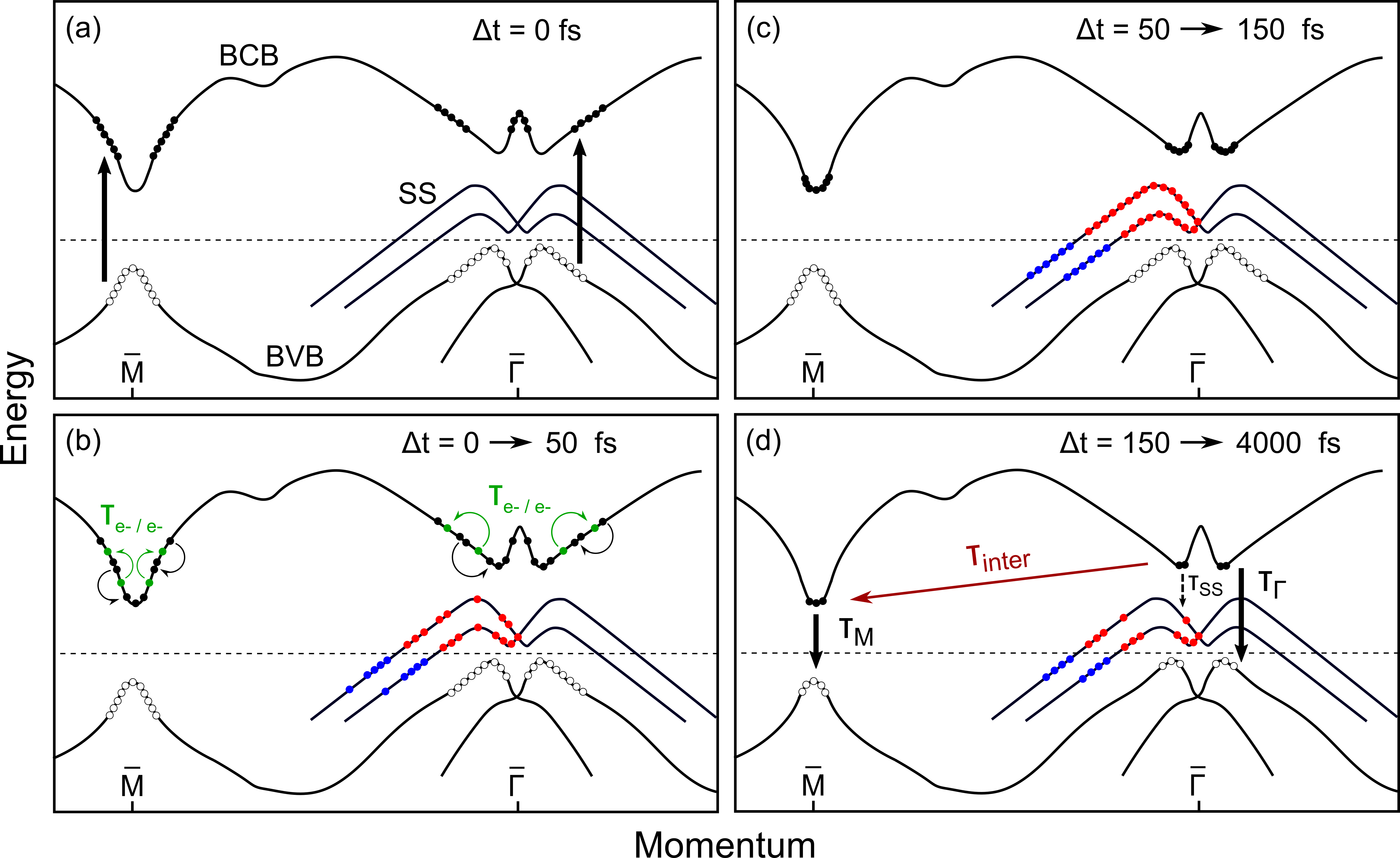}
\caption{\textbf{Cartoon of the transitions and scattering processes at the $\bar{\Gamma}$ and $\bar{M}$ points of the BZ.} (a) Pump induced optical transition from the top of the bulk valence band (VB) to the bulk conduction band (CB) both at the $\bar{\Gamma}$ and $\bar{M}$ points of the BZ. Electrons and holes associated to this process are represented by full and empty black balls, respectively. (b) Scattering of electrons in the bulk CB and  thermalization of the electrons in the surface state (SS) leading to a  transient population of the SS close to the $\bar{\Gamma}$ point (as represented by red (electrons) and blue (holes) balls). Additional electron-electron scattering processes mentioned in the main text from electrons scattered to higher energy are represented by green arrows. Note that this mechanism occurs over at least 1 ps according to time traces in Fig. \ref{fig4}(h,j) but it is only represented on panel (b) of the cartoon for clarity. (c) Accumulation of electrons at the minimum of the CB both at $\bar{\Gamma}$ and $\bar{M}$ points and maximized thermally assisted population of SS. (d) Relaxation of electrons from the bottom of the CB to the top of the VB through different inter- and intra- bands channels at $\bar{\Gamma}$ and $\bar{M}$ points.}
\label{fig5}
\end{figure*}

In Fig. \ref{fig4}(f-h), we look at the temporal dynamics of the electrons in the conduction states close to $\bar{\Gamma}$. To do so, we plot the time evolution of the photoemission intensity integrated in certain energy/momentum windows. We focus on three boxes in the BCB (cyan, blue and pink) and three smaller boxes in the $SS$ (red, green, yellow). Panel (g) shows the corresponding time traces for the $SS$: the three curves show a superimposed leading edge corresponding to a simultaneous population at different energy/momentum locations, as well as a common maximum at $\Delta t \approx 180$ fs (see inset). Then, the intensity decreases with a characteristic exponential relaxation, which is faster when the energy of the integration box is larger ($408 \pm 60$ fs for the red, $583 \pm 60$ fs for the green and $1050 \pm 60$ fs for the yellow boxes). These different features are fully compatible with a thermal population due to a transient increase of the electronic temperature of the system and associated Fermi edge broadening in the first hundreds fs after photoexcitation (see supplementary material of our recent publication\cite{kremer2022}). Similar dynamics has been observed using $1.5$ eV pump and $6$ eV probe\cite{clark2022ultrafast}. Due to the low heat capacity of the electrons, the electronic temperature can reach up to $ 1100$ K in \hbox{$\alpha$-GeTe(111)}\cite{kremer2022}.  We now focus on the electron dynamics in the BCB as described by the cyan, blue and pink regions in Fig. \ref{fig4}(h). The first and the second show a similar behaviour, namely a quasi-instantaneous population (within the duration of the cross correlation of the pump and probe pulses)  which is a signature of a direct optical transition from the BVB to the BCB, and a fast relaxation with a sub-$50$ fs exponential relaxation. The pink time trace is different since it exhibits (i) a delayed leading edge and maximum that we can interpret as a population of electrons coming from states with higher energy, and (ii) a slower relaxation time of $845 \pm 45$ fs.

The same analysis is now done for the BCB at the $\bar{M}$ point using the MM. In Fig. \ref{fig4}(i) is shown the ARPES dispersion of the BCB close to the $\bar{M}$ point with a series of pertinent integration boxes. The corresponding time traces are plotted in panel (j). Green and purple curves are characteristic of an optical direct excitation of carriers from the BVB to the BCB with a fast relaxation time of \hbox{$29 \pm 5$ fs} and $90 \pm 10$ fs respectively.  As previously observed at $\bar{\Gamma}$, at the bottom of the BCB at $\bar{M}$ (brown line), this time is two orders of magnitude bigger because the electrons accumulate and take time to recombine ( $1750 \pm 100$ fs). The leading edge positions of each of these three time traces are different : this is again a signature of the delayed population of the BCB in different energy regions. It is important to further notice that both at $\bar{\Gamma}$ and $\bar{M}$ points, the relaxation dynamics of high energy electrons in the BCB cannot be fitted by a single exponential (the fast one we discussed) but another exponential relaxation is needed. Its relaxation constant is much longer than the pump pulse duration. We conclude that this slow electron dynamics must be coming from electrons scattered to higher energy by electron-electron scattering processes\cite{williams2017direct}. A high transient electronic temperature for electrons in the BCB could also give rise to an electronic population persisting on a $500$ fs timescale. However, we discard this hypothesis, since their relaxation constant does not vary much with energy (see Fig. \ref{fig4}(h), contrarily to the case of electrons in the SS (see Fig. \ref{fig4}(g)). Finally, in Fig. \ref{fig4}(k) we compare the electron dynamics at the bottom of the BCB at $\bar{\Gamma}$ and $\bar{M}$ points. In both cases, the leading edge and consequently the population time is the same, but the relaxation time is different. It respectively equals $860 \pm 100$ fs and $1750 \pm 100$ fs. We explain this difference to be likely due to additional scattering channels from the BCB at the $\bar{\Gamma}$ point i.e. (i) BCB to $SS$ and, (ii) BCB from $\bar{\Gamma}$ to $\bar{M}$ due to higher relative energy position as above discussed.

In Fig. \ref{fig5}, we give a schematic representation of the out-of-equilibrium dynamics of the electronic band structure and pertinent time regimes of the carriers dynamics in the full BZ. (a) At time zero ($\Delta t = 0$ fs), electrons (full balls) are optically excited from the BVB to high energy location in the BCB by receiving $1.55$ eV energy from the pump pulse, both in the vicinity of $\bar{\Gamma}$ and $\bar{M}$ points, as schematized by vertical thick black lines. Corresponding holes (empty balls) are created in the BVB. The system is brought to a highly non-equilibrium state. (b) Excited electrons in the BCB lose energy by intra-band scattering processes at both high-symmetry points ($\Delta t = 0 \rightarrow 50$ fs) and the $SS$'s intensity increases above $E_{F}$ (red balls) whereas it decreases below (blue balls). It is explained by the fact that electron-electron and electron-phonon scattering acts to bring the system away from the highly non-equilibrium state towards a thermal distribution, typically with a high electronic temperature. The lattice temperature increases as well, but to a lesser extent due to its larger heat capacity. (c) This is completed in the first few hundreds fs ($\Delta t = 50 \rightarrow 150$ fs) with electrons accumulating at the bottom of the BCB both at $\bar{\Gamma}$ and $\bar{M}$ points, and a maximum of the electronic temperature associated to a broad Fermi edge and largest $SS$ population above $E_{F}$. (d) After about $150$ fs the electronic temperature decreases by transfer of energy to the lattice sub-system  and the electronic population in the SS converges towards the Fermi level. The system is going back to the equilibrium state with recombination of electrons and holes from the BCB to the BVB, with characteristic relaxation times at different high symmetry points (see vertical solid  lines). Asymmetry in the relaxation times can likely be explained by additional scattering channels between BCB and $SS$ close to $\bar{\Gamma}$ point (see dashed vertical lines), as observed in the case of the BVB\cite{clark2022ultrafast}. The progressive reduction of the electronic temperature causes the reduction of the SS's population above $E_{F}$.  

In addition to these transient electronic population dynamics, a coherent motion of lattice modes can be triggered by the nearly-instantaneous perturbation through optical excitation as well, with an associated time evolution of the electronic band structure. As previously reported, this is observed in the case of \hbox{$\alpha$-GeTe(111)} with a particular delayed displacive excitation of a coherent phonon  mechanism assisted by the creation of a transient surface photovoltage, leading to a field-induced and time dependent Rashba coupling in the material\cite{kremer2022}.


\begin{center}
\textbf{IV. CONCLUSIONS}
\end{center}

To summarize, the band structure and the ultrafast dynamics of FE \hbox{$\alpha$-GeTe(111)} have been experimentally explored in the full BZ thanks to state-of-the-art tr-ARPES. We report the full electronic band structure both in the valence and conduction states, in very good agreement with BSF calculations. Valence states dispersions are in line with previous ARPES measurements and our exploration of the conduction states gives access to the dispersion above $E_{F}$ which has not been reported to date, excepted in a region very close to the center of the BZ and to $E_{F}$. In particular, an upper limit for the electronic band gap of 0.85 eV is found, confirming its indirect nature in agreement with DFT calculations. In addition, we give insights into the out-of-equilibrium dynamics with a discussion of the transitions and scattering processes occurring after excitation by fs light pulses. The observed temporal dynamics are typical of a semiconducting material where the carriers excess energy is relaxed by scattering events which transiently populate bands at different locations in the BZ and finally accumulate at the CB minima. Characteristic time constants are extracted and confirm the interplay of electron-electron and electron-phonon scattering events. We also observe a Rashba split $SS$'s population mediated by a transient evolution of the electronic temperature in the first ps after excitation. Overall, our study brings valuable informations on the electronic band structure of \hbox{$\alpha$-GeTe(111)}, knowledge of which is important for understanding the fundamental properties of the material, in particular in the perspective of its use in modern electronics, and ultrafast spintronics as a FERSC.

\bigskip


\begin{center}
\textbf{ACKNOWLEDGMENTS}
\end{center}

G. Kremer, C. Nicholson and C. Monney acknowledge support from the Swiss National Science Foundation  Grant \hbox{No. P00P2$\_$170597}. This work was supported by the Max Planck Society, the European Research Council (ERC) under the European Union’s Horizon 2020 research and innovation program (Grant No. ERC-2015-CoG-682843), and the German Research Foundation (DFG) within the Emmy Noether program (Grant No. RE3977/1). J. Minár and L. Nicolaï would like to thank the project QM4ST with reg. \hbox{No. CZ.02.01.01/00/22$\_$008/0004572}, co-funded by the ERDF as part of the MŠMT and TWISTnSHINE, funded as project No. LL2314 by Programme ERC CZ. G. Springholz acknowledges support by the Austrian Science Funds,
Grant No. 10.55776/PIN6540324 and I-4493, and JKU-Linz Grant 296/LIT-2022-11-SEE-131.


\bibliography{biblio} 
\end{document}